\documentclass[sn-mathphys,Numbered]{sn-jnl}% Math and Physical Sciences Reference Style
\usepackage{graphicx}%
\usepackage{multirow}%
\usepackage{amsmath,amssymb,amsfonts}%
\usepackage{amsthm}%
\usepackage{mathrsfs}%
\usepackage[title]{appendix}%
\usepackage{xcolor}%
\usepackage{textcomp}%
\usepackage{manyfoot}%
\usepackage{booktabs}%
\usepackage{algorithm}%
\usepackage{algorithmicx}%
\usepackage{algpseudocode}%
\usepackage{listings}%

\usepackage{float}
\usepackage{hyperref}
\usepackage{cleveref}
\usepackage{siunitx}

\usepackage{textcomp}

\DeclareSIUnit[quantity-product = { }] \debye{\text{D}}
\theoremstyle{thmstyleone}%
\theoremstyle{thmstyletwo}%

\theoremstyle{thmstylethree}%

\begin{document}

\title[Fluid fibres in  true 3D ferroelectric liquids]{Fluid fibres in  true 3D ferroelectric liquids}


\author[1]{\fnm{Alexander} \sur{Jarosik}}%\email{alexander.jarosik@ovgu.de}

\author[1]{\fnm{Hajnalka} \sur{N\'adasi}}
%\email{hajnalka.nadasi@ovgu.de}

\author[2]{\fnm{Michael} \sur{Schwidder}}%\email{hajnalka.nadasi@ovgu.de}

\author[3]{\fnm{Atsutaka} \sur{Manabe}}
\author[4]{\fnm{Matthias} \sur{Bremer}}

\author[4]{\fnm{Melanie} \sur{Klasen-Memmer}}%\email{melanie.klasen-memmer@merckgroup.com}
\equalcont{These authors contributed equally to this work.}

\author*[1]{\fnm{Alexey} \sur{Eremin}}%\email{alexey.eremin@ovgu.de}
\equalcont{These authors contributed equally to this work.}

\affil*[1]{\orgdiv{Department Nonlinear Phenomena}, \orgname{Institute of Physics, Otto von Guericke University}, \orgaddress{\street{Universitaetsplatz 2}, \city{Magdeburg}, \postcode{39106},  \country{Germany}}}
\affil[2]{\orgdiv{Department Industrial Chemistry}, \orgname{Institute of Chemistry, Otto von Guericke University}, \orgaddress{\street{Universitaetsplatz 2}, \city{Magdeburg}, \postcode{39106},  \country{Germany}}}
\affil[3]{\orgdiv{Independent researcher} \city{Bensheim},  \country{Germany}}
\affil[4]{\orgdiv{Merck Electronics KGaA,} \city{Darmstadt},  \country{Germany}}


\abstract{We demonstrate an exceptional ability of a high-polarisation 3D ferroelectric liquid to form freely-suspended fluid fibres at room temperature. Unlike fluid threads in modulated smectics and columnar phases, where translational order is a prerequisite for forming liquid fibres,  recently discovered ferroelectric nematic forms fibres with solely orientational molecular order. Additional stabilisation mechanisms based on the polar nature of the mesophase are required for this. We propose a model for such a mechanism and show that these fibres demonstrate an exceptional non-linear optical response and exhibit electric field-driven instabilities.}




\maketitle

\section{Introduction}\label{sec1}
The phenomenon of freely suspended fluid fibres in non-Newtonian fluids is a fascinating topic with many practical applications. From living organisms to technological applications, such as spider silk, polymeric melts, textiles, and photonic devices, this phenomenon is observed in a wide range of areas~\cite{Vollrath2001,Kerkam:1991ji, Wang.2023}. Fibre-based materials are particularly desirable for wearable electronics due to their flexibility and stretchability, providing advantages over solid-state materials for the development of the next generation of electronic devices~\cite{Stoppa.2014, Heo.2018}. Recent developments in smart materials and wearable design have created opportunities for the creation of structured and multifunctional materials, resulting in further research in this field.

Fibres can result from solidified liquids or glasses, which initially start as a liquid~\cite{Mercader.2010,Sparkes.2019}. In Newtonian liquids, long liquid filaments cannot form due to the Rayleigh-Plateau instability~\cite{Rayleigh:1879vw}. However, non-Newtonian fluids such as polymeric solutions and melts can form cylindrical filaments and jets during the process of thinning fluid bridges suspended between two supports or during droplet detachment~\cite{Goldin.1969jvb}.

The capillary-induced thinning of liquid filaments has been demonstrated in multiple studies as an effective method for characterising rheological materials~\cite{CLASEN.2006,Anna.2001}. This approach can be utilised as a rheometric device for characterising rheological materials, which enables the determination of their flow and deformation properties under different conditions~\cite{Anna.2001}. The usefulness of capillary-induced thinning of liquid filaments for rheological material characterisation and as a rheometric device has been demonstrated in several studies, making it a valuable tool in the field of materials science.

 However, structured fluids, such as some types of liquid crystals, can form stable fibres~\cite{Mahajan:1999ua,Cheong2001,Cheong2002}. 
Experiments with pure thermotropic liquid crystals have shown that the orientational anisotropy of nematics alone does not affect filament stability~\cite{Mahajan:1999ua} . Columnar~\cite{Gharbia.1990,Fontes:1988vl,Safinya:1985io,Kamenskii:1983uq,VanWinkle1982} or modulated smectic order~\cite{Baily2007,Jakli2003,Eremin:2005by,Eremin:2012gg,TambaMG:2015ga} is the prerequisite for stabilising the filament structure.

Nematics are liquids exhibiting orientational molecular order and finding various applications from displays to electro-optics, photonics and sensorics~\cite{deGennes:1995vg}. Most nematics show quadrupolar uniaxial and rarely biaxial types of order. Possibility to stabilise the nematic phase assuming dipolar correlations were proposed by Max Born as early as 1916~\cite{Born.2016}.
 The symmetry of such a liquid corresponds to that of the ferroelectric nematic. The dipolar symmetry of the N$_{\rm{F}}$ phase sharply contrasts the quadrupolar symmetry of common nematics. It is accompanied by breaking the head-tail invariance of the nematic director. Although such symmetry breaking is quite common at interfaces, the experimental realisations of the bulk ferroelectricity in nematics remained elusive. A few polymeric and lyotropic systems showed indications of polar order in bulk. Only in 2017, Nishikawa et al. synthesised a thermotropic liquid crystal with a 1,3-dioxane unit in the mesogenic core (DIO) and exhibiting two nematic phases with a low-temperature phase distinguished by the spontaneous polarisation aligned with the director~\cite{Nishikawa.2017}. Around the same time, Mandle et al. designed a series of mesogens also exhibiting polymorphic nematic phases with a low-temperature ferroelectric nematic phase~\cite{Mandle.2017.nematic-nematic,Mertelj.2018}.
 
Ferroelectric nematics are distinguished by an unusually high spontaneous polarisation typically within the range of a few \textmu C~cm$^{-2}$~\cite{Lavrentovich.2020,Chen.2020,Mandle.2021,Brown.2021ysv,Sebastian.2021}. Mertelj and Sebastian have demonstrated that the transition into the N$_{\rm{F}}$ phase involves a significant softening of splay deformations and an increase in polar correlations~\cite{Sebastian.2020}. Consequently, electrostatic interactions play a very important role in defining the materials' macroscopic properties. Avoiding the polarisation splay results in stiffening the splay elasticity~\cite{Zavvou.2022gmk}. Another necessary consequence of the polar symmetry in the nematic phase is the possibility of developing spontaneous splay, as theoretically investigated by Pleiner and Brand~\cite{Pleiner.1989}. Although the field stabilised N$_{\rm{F}}$ thread-like structures were found by Nishimura et al. \cite{Nishimura.2022} in experiments on electrostatic actuators, spontaneous formation of fibres remained unexplored. 

In this paper, we experimentally demonstrate that ferroelectric nematics spontanously form liquid filaments. The filament formation occurs regardless of whether the compound undergoes a direct isotropic-N$_{\rm{F}}$ or nematic-N$_{\rm{F}}$ transition. These filaments can be created by mechanically pulling from a droplet or through the electro-capillary instability of a droplet in a vertical electric field. Using non-linear optical microscopy, we show that those filaments exhibit a remarkably efficient optical second harmonic generation. A model based on the polar structure of the phase is proposed to explain filament formation.

\section{Results}\label{sec2}
\subsection{Filament stability and optical properties}
\begin{figure}[t]
\centering
\includegraphics[width=\columnwidth]{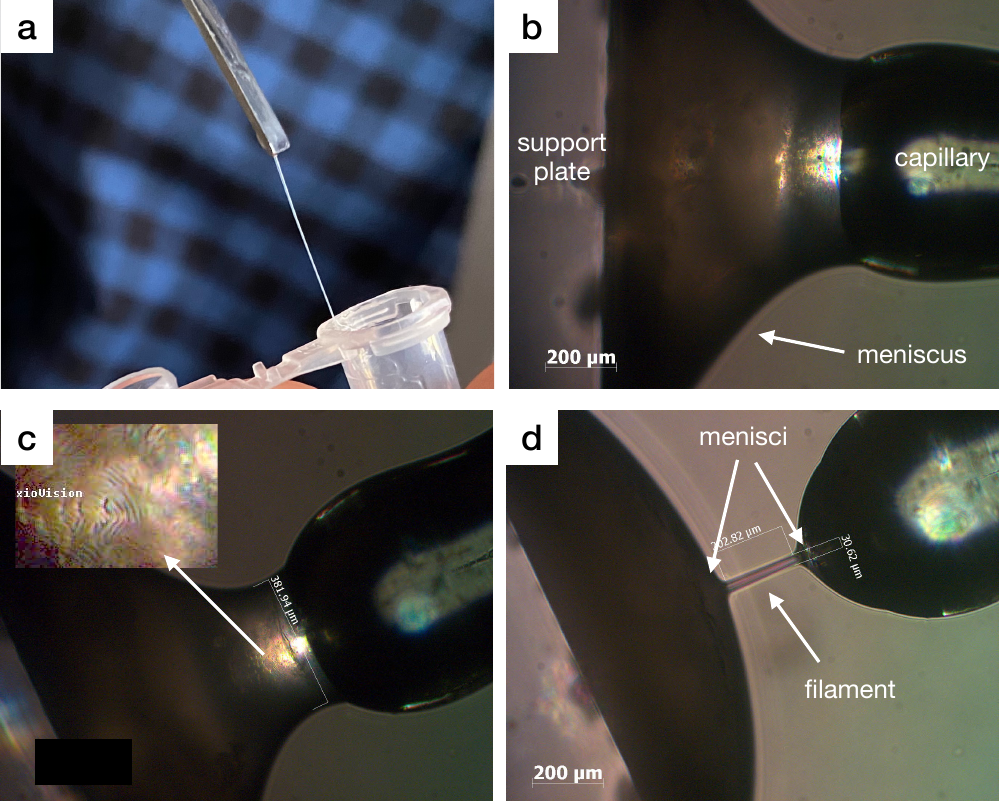}
\caption{Filament formation in Material \textbf{2}: \textbf{a} Spontaneously formed freely-suspended filament pulled at room temperature using a spatula. Short birefringent bridges can be formed below the Rayleigh-Plateau limit in the nematic \textbf{b} and the M phases \textbf{c}; Long filaments appear in the N$_{\rm{F}}$ phase \textbf{d} }\label{fig:bridges}
\end{figure}

A striking feature of pulling filaments in the ferroelectric nematic becomes apparent when one is trying to pick a small amount of liquid crystal with a spatula from a vial. Long and occasionally, multiple fibres become attached to the spatula (Fig.~\ref{fig:bridges}a).
\begin{figure}[ht]
\centering
\includegraphics[width=0.9\columnwidth]{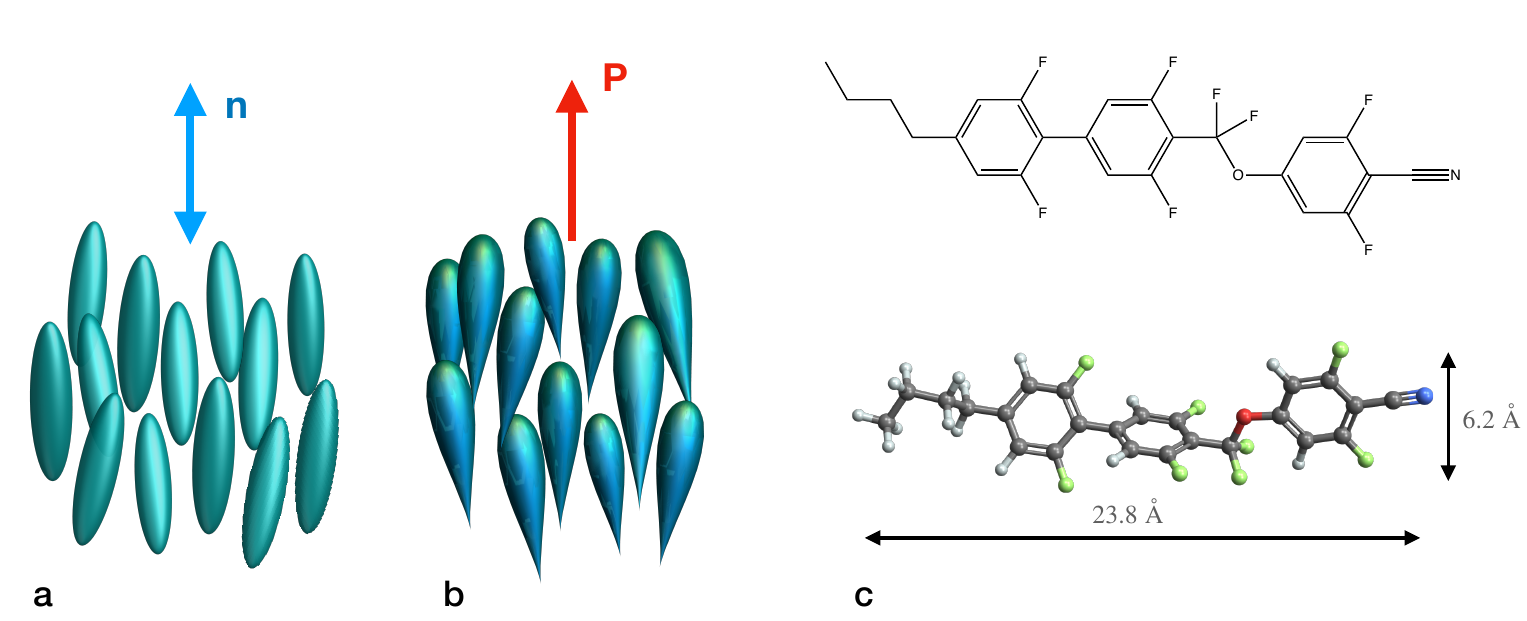}
\caption{\textbf{a} A sketch of a nematic phase with the quadrupolar order; \textbf{b} Molecular arrangement in the ferroelectric nematic phase (N$_{\rm{F}}$) with the polar order; \textbf{c} Chemical formula of compound \textbf{1} with a 3D structure.}\label{fig:molecule}
\end{figure}
We explored this behaviour in two materials: compound \textbf{1} is a single-component mesogen exhibiting a direct transition from the isotropic to the ferroelectric nematic (N$_{\rm{F}}$) phase (Fig.~\ref{fig:molecule}):
\\

isotropic (\qty{19.6}{\celsius} N$_{\rm{F}}$) \qty{44.0}{\celsius} crystal,\\

\noindent and material \textbf{2}, a mixture exhibiting the transitions
\\\ 

isotropic \qty{87.0}{\celsius} N \qty{57.0}{\celsius} M \qty{46.0}{\celsius} N$_{\rm{F}}$ \qty{-43.0}{\celsius} crystal.\\

Both compounds exhibit the room temperature N$_{\rm{F}}$ with a high spontaneous polarisation in the rang of 5 - \qty{6}{\micro\coulomb\per\cm\squared}. The mesogen in compound \textbf{1} has a strong dipole moment $\mu=\qty{11.3}{\debye}$, and its properties are described in~\cite{Zavvou.2022gmk,Manabe.2021}.

A drop of isotropic liquid spanned between two supports forms a catenoid-shaped profile of a liquid bridge formed by two menisci. Surface tension and the wetting conditions at the support determine its shape. As the distance between the supports increases, the bridge collapses due to the Plateau-Rayleigh instability. 
The bridge becomes birefringent when cooling to the nematic phase N (Fig.~\ref{fig:bridges}b). A grainy character of the microscopic texture observed between crossed polarisers suggests a disordered polydomain arrangement in the absence of the alignment within the meniscus. A slightly better alignment was observed in the central part of the bridge. Strong fluctuations of the nematic director result in the scattering of light observed as a typical flickering of the microscopic texture. Those fluctuations become quenched upon the transition into the intermediate M phase (Fig.~\ref{fig:bridges}c). The microscopic texture remains grainy in the meniscus and partially aligned in the central part of the bridge. Labyrinthine structures appear at the surface of the fluid bridge (Fig.~\ref{fig:bridges}c).
The length of bridges in the N and M phases is not greater than that in the isotropic phase.

When entering the ferroelectric nematic phase N$_{\rm{F}}$, the situation changes drastically. The director fluctuations become visible again. Stretching the bridge above the critical length does not lead to immediate collapse. Rather, a cylindrical filament will form between the pair of separated menisci (Fig.~\ref{fig:bridges}d).
%The filament can be stretched up to the length of XX\qty{11}{\micro\metre}. 
Dust particles occasionally trapped at the filament surface help to visualise the material flow at the surface. The particles move slowly during the filament's extension, now and then exhibiting slow circulating motion around the filament axis. These observations suggest that there is no significant flow present at the filament surface in the steady state. 

Optical anisotropy indicates anisotropic molecular order, and most liquid crystals exhibit birefringence. In nematics, the birefringence is connected to the orientational order parameter. In the N$_{\rm{F}}$ filaments, the slow optical axis (with the largest refractive index) points along the filament axis, as determined using a variable retarder. The experiments on aligned samples in planar cells showed that the same axis is parallel to the director in the N and the N$_{\rm{F}}$ phases. This suggests that the molecular orientation in the filament is parallel to the filament axis. This phenomenon is akin to the case of spider silk formed by the hardening of silk fibres through the extensional flow. 

\subsection{Behaviour in electric field}
\begin{figure}[ht!]
\centering
\includegraphics[width=\columnwidth]{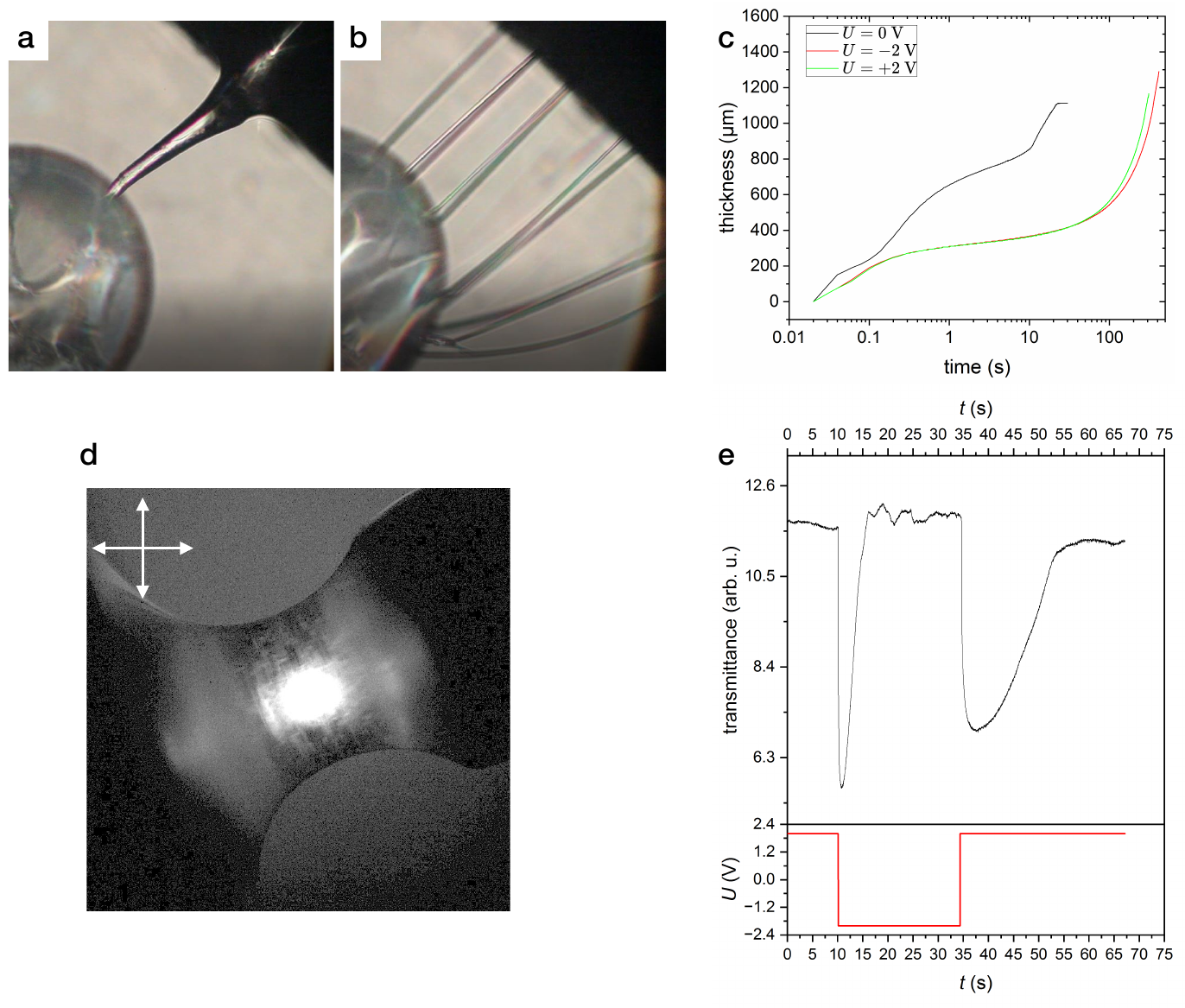}
\caption{Behaviour of fluid bridges and filaments in an external electric field: \textbf{a} Electrically stabilised filament becomes destabilised on the inversion of the field polarity when the new bunches of filaments appear \textbf{b}. \textbf{c}: Stabilising effect of the electric field on the thinning dynamics of filaments. \textbf{d}: A thick LC bridge observed between crossed polarisers exhibit electrooptical switching manifested by the field-dependence of the optical transmittance shown in \textbf{e}.
}\label{fig:e_field}
\end{figure}

Spontaneous polarisation in the nematic phase gives rise to the whole zoo of remarkable effects in an electric field, such as explosive electrostatic instabilities in droplets \cite{Barboza.2022}, formation of dendritic structure in droplets \cite{Mathe.2023}, and behaviour of soliton walls in N$_\textrm{F}$ \cite{Basnet.2022bek}, light-driven propulsion of ferroelectric droplets \cite{Marni.2023}.
Fluid filaments also show remarkable behaviour in an external electric field applied axially along the filament axis. Thick bridges can switch between electrical and optical states when the field is reversed within a moderate range (as shown in Fig.~\ref{fig:e_field} d,e). To demonstrate this, we can measure the optical transmittance of a bridge between crossed polarisers as the field is varied. When the voltage is reversed, the transmittance transiently decreases and increases again. However, if the same voltage is applied again after the field is removed, there is only a negligible change in transmittance. 

In high fields, in the case of non-isolated electrodes, convective patterns occur, and the nematic alignment becomes disturbed.

Optical switching is not observed in thin filaments, but the electric field can have stabilising or destabilising effects depending on polarity. In the stabilising case, stronger fields can slow down or reverse thinning dynamics through the supply of material from the meniscus (Fig.~\ref{fig:e_field}c). However, field reversal can cause the collapse of the filament. Asymmetric metal/glass interfaces can lead to a destabilising field after the collapse of the filament, resulting in a bursting instability and a cluster of sideways-injected filaments (Fig.~\ref{fig:e_field}a,b). 

However, if the filament is pulled from a droplet in an electric field, the field effect has a stabilising character independent of the polarity. As shown in Fig.~\ref{fig:e_field}c, applying the field as low as \qty{2}{\volt\per\micro\meter} increases the thinning time by order of magnitude compared to the field-free case.

\subsection{Nonlinear optical behaviour}
Ferroelectric nematics exhibit an exceptionally high second harmonic generation efficiency (SHG)~\cite{Folcia.2022,Nishikawa.2021,Li.2021}. This is a direct consequence of the polar mesophase symmetry and high values of the second-order molecular hyperpolarisability responsible for the conversion of the infrared light ($\lambda=\qty{880}{\nano\meter}$) of the primary beam to the second harmonic, SH ($\lambda=\qty{440}{\nano\meter}$). As a result, the filaments prepared in the N$_{\rm{F}}$ exhibit strong SH generation.

\begin{figure}[t]
\centering
\includegraphics[width=\columnwidth]{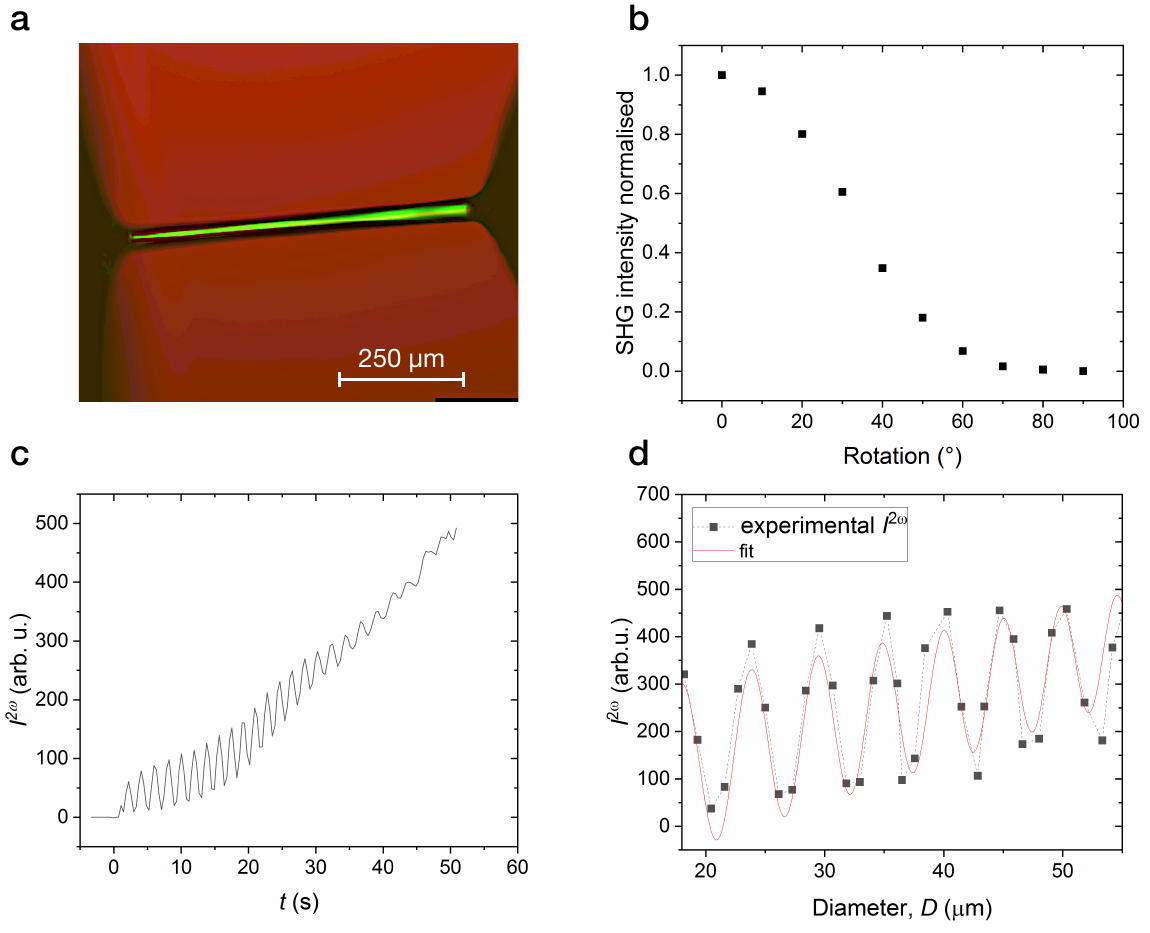}
\caption{Nonlinear optical behaviour of the ferroelectric fibres: \textbf{a} SHG microscopy image superimposed with the brigt field image of the filament at $T=\qty{21}{\celsius}$; \textbf{b} Angular dependence of the SHG signal $I^{2\omega}$; \textbf{c} Temperature dependence of the SHG signal of a rapidly thinning filament. \textbf{d} Maker fringes in the thickness dependence of a filament. The red curve is a fit with Eq.~\ref{eq:shg_fit2}.
}\label{fig:shg}
\end{figure}

Fig.~\ref{fig:shg}a exemplifies an image of a filament captured by SHG-microscopy. The record of the angular dependence of the SH signal shows that the highest signal is obtained when the polarisation of the primary laser beam is parallel to the filament axis (Fig.~\ref{fig:shg}b), suggesting that the polar axis is aligned parallel to the nematic director in the N$_{\rm{F}}$ phase.  According to the Kleinman principle, $C_{\infty h}$ symmetry of the N$_{\rm{F}}$ phase results in the second-order polarisation: 
\begin{equation}
  \mathbf{P}_{2 \omega}=\varepsilon_0 \left[\begin{array}{cccccc}
0 & 0 & 0 & 0 & d_{31} & 0 \\
0 & 0 & 0 & d_{31} & 0 & 0 \\
d_{31} & d_{31} & d_{33} & 0 & 0 & 0
\end{array}\right]\left[\begin{array}{c}
\left(E_{\omega x}^S\right)^2 \\
\left(E_{\omega y}^S\right)^2 \\
\left(E_{\omega z}^S\right)^2 \\
2 E_{\omega y}^S E_{\omega z}^S \\
2 E_{\omega z}^S E_{\omega x}^S \\
2 E_{\omega x}^S E_{\omega y}^S
\end{array}\right],
\end{equation}

\noindent where $d_{31}$ and $d_{33}$ are the second-order nonlinear optical coefficients in a coordinate frame with axis "3" (Z) directed along the polar axis of the nematic. Thus, the fundamental light polarised along the $X$ axis (extraordinary) of a thin N$_{\rm{F}}$ slab results in the SHG determined by $d_{33}$. The light polarised orthogonal to the $X$ axis generates an SH signal determined by $d_{31}$. 

\begin{figure}[t]
\centering
\includegraphics[width=0.9\columnwidth]{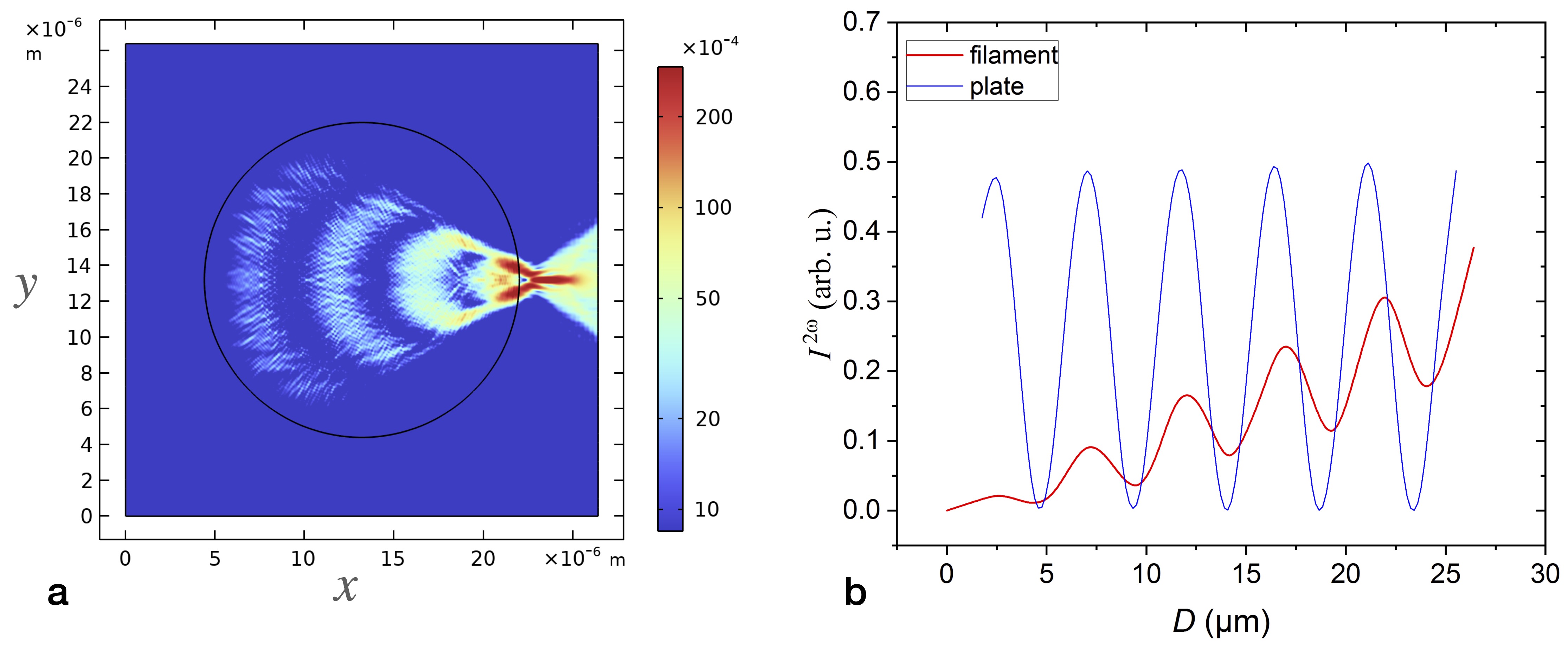}
\caption{Numerical simulation of SHG in a filament: \textbf{a} Average power of SH in a filament with a radius $r=10\lambda$, average refractive index $n=1.6$, and $\Delta n_{\mathrm{d}}=0.093$.\textbf{b} SHG intensity plotted as a function of thickness (diameter) for a filament (red curve) and a slab (blue curve) with identical nonlinear optical coefficients.
 }\label{fig:shg_model}
\end{figure}

As the thickness of the slab is varied, the $I^{2\omega}$ exhibits a typical periodic dependence called Maker fringes. This occurs due to the energy exchange between the primary and the SH beams (see also Fig.~\ref{fig:shg_model}b).

In the case of a plane parallel slab, the Maker fringes can be described by the equation:
\begin{equation}
  I^{2\omega}(D)=C d_{\mathrm{eff}}^2 I^{\omega}\frac{\sin^2{\big(\frac{2 \pi}{\lambda} \Delta n_\mathrm{d} L\big)}}{\big(\frac{2 \pi}{\lambda} \Delta n_\mathrm{d}\big)^2},
  \label{eq:shg_fit}
\end{equation}

\noindent where the coefficient $C$ is determined by the Fresnel transmission coefficients for the fundamental, the SH lights, $\lambda$ is the wavelength, and $L$ is the sample thickness, $d_{\mathrm{eff}}$ is the effective NLO coefficient, and $\Delta n_{\mathrm{d}}=n^{2\omega}-n^{\omega}$ is the dispersion. Thus, analysis of Maker fringes allows us to estimate the LC material's second-order nonlinear optical (NLO) coefficients of the LC material~\cite{Vivek.2021,Okamoto.1992,Jerphagnon.1970}.

To determine the NLO coefficients, we can use the periodicity of the fringe pattern to calculate the difference in refractive indices, even if we don't know the exact values of $n^{2\omega}$ and $n^{\omega}$. By comparing the maximal intensities to those of a reference like $\alpha$-Quartz, we can extract the NLO coefficients for the material under study.
In the case of filaments, the thickness dependence of $I^{2\omega}$ can be recorded during filament thinning. However, the fringe pattern is different from that of the plane parallel slab. The periodic fringe pattern is superimposed with a baseline growing with increasing filament diameter. In the range of rather small diameters $D$, the fringe pattern can be approximated by the function
\begin{equation}
  I^{2\omega}(D) = C_{\textrm{base}}D+C_{\textrm{A}}\sin^2{\big(\frac{2 \pi}{\lambda} \Delta n_\mathrm{d} D\big)},
   \label{eq:shg_fit2}
\end{equation}
\noindent where $C_{\textrm{base}}$ and $C_{\textrm{A}}$ are fit parameters for the baseline and the amplitude, respectively.

The difference occurs due to the circular shape of the filament cross-section allowing the rays with different path lengths to contribute to the net SH power. A simulation of the SH generation in a filament using the finite element method is shown in Fig.~\ref{fig:shg_model} qualitatively confirms this behaviour. The net power exhibits fringes within the filament cross-section as shown in Fig.~\ref{fig:shg_model}a. 
%Comparing the SHG behaviour of a filament and a slab allows us to roughly estimate the nonlinear optical coefficient $d_{33}$.
 
%A rough estimation of the radius dependence of the net power can be given by 
%
%\begin{equation}
%  XX
%\end{equation}
%\noindent where $R$ is the filament radius. 
Even in this case, the period of the fringe pattern equals  $\lambda/\Delta n_{\mathrm{d}}$. This allows us to determine $\Delta n_{\mathrm{d}}=0.093$ from the thickness dependence of the $I^{2\omega}(D)$. 
The slope of the baseline allows us to roughly estimate the nonlinear optical coefficient $d_{33,\textrm{F}}$ of the filament material. The simulation in Fig.~\ref{fig:shg_model}b compares the thickness dependences of the SHG signal of a filament and a reference slab with identical coefficients $d_{33}$. Taking the mean intensity of the slab as a reference, the baseline of the normalised filament intensity is proportional to the diameter $D$, $I^{2\omega}_{\textrm{fil}}=\kappa D$, where $\kappa\approx \qty{1e-9}{\micro\meter^{-1}}$. The same approach can be applied to the experimental data giving the ratio
\begin{equation}
  \frac{\langle I^{2\omega}_{\textrm{F}}\rangle}{\langle I^{2\omega}_{\textrm{quartz}}\rangle}= \frac{d_{33,\textrm{F}}^2}{d_{33,\textrm{Q}}^2}\kappa D,
\end{equation}
 
\noindent where $d_{33,\textrm{Q}}= \qty{0.4}{\pico\meter\per\volt}$ is the nonlinear optical coefficient of $\alpha$-Quartz. The slope of the baseline determined in the experiment is $\approx \qty{9}{\per\micro\meter}$ giving an estimation $d_{33}(\textrm{F})\approx \qty{2.2}{\pico\meter\per\volt}$.

This value is more than five times higher than that of $\alpha$-Quartz, making this material stand out. At the same time, it is smaller than the values reported by Folcia et al.~\cite{Folcia.2022} for another ferroelectric nematogen mesogen, RM734, aligned in an electric field. This difference can be attributed to the different chemical structures of our compounds containing several fluorine substituents.
\section{Discussion}
Polar, SHG-active filaments were initially discovered in the polarisation-modulated smectic phase of bent-core liquid crystals~\cite{Jakli2003,Eremin:2012gg}. These filaments display exceptional stability for a few days. In contrast, the stability range of the nematic filaments is restricted to 10 - 30 seconds in compound \textbf{1} to several minutes in material \textbf{2}. The filaments thin retaining their cylindrical shape until they eventually break. What is the reason for the stability of the nematic filaments? 

\begin{figure}[t]
\centering
\includegraphics[width=\columnwidth]{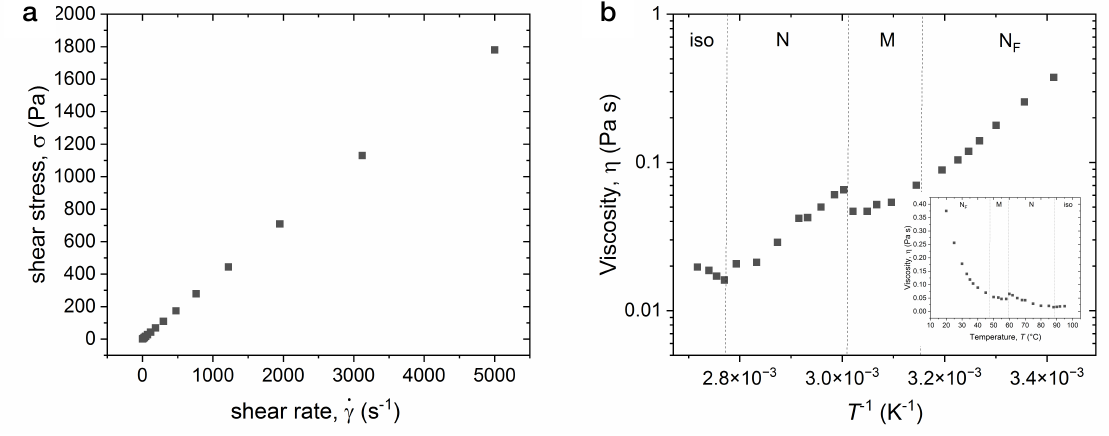}
\caption{Rheological behavioiur of compound \textbf{1}: \textbf{a}
flow curve measured at $T=\qty{40}{\celsius}$, \textbf{b} Temperature dependence of the viscosity.}
\label{fig:rheo}
\end{figure}

In polymeric systems, it is the nonlinear rheology responsible for filament formation. 
Intensive theoretical and experimental studies of the filament formation and jet break-up showed that extensional rheological response dictates whether or not a stable filament is formed~\cite{Goldin.1969jvb,Bogy.1979,Eggers.1997}. During the necking process, as a fluid bridge is thinning, elastic tensile stresses resist the pinching driven by the capillary forces~\cite{Anna.2001}. 
In our case, however, the mesogens are low-weight molecules. In the whole temperature range of the liquid crystal phases, the rheological behaviour is Newtonian with a linear flow curve (Fig.~\ref{fig:rheo}a), excluding this mechanism of filament stabilisation. 
The viscosity exhibits Arrhenius dependence in the conventional and ferroelectric nematic phases (N and N$_{\rm{F}}$). Non-Arrhenius behaviour occurs in the intermediate M phase.
It appears that nematic filaments occur only in the ferroelectric nematic phase regardless the presence of the mesophase above N$_{\rm{F}}$. Same behaviour is observed in compounds with the direct iso-N$_{\rm{F}}$ and iso--N--M--N$_{\rm{F}}$ transitions. These observations suggest that ferroelectric polarisation is essential for filament stability. Indeed the instability of a fluid cylinder occurs due to the growth of surface distortion modes when the aspect ratio of the cylinder exceeds a critical value. Those distortions are accompanied by the decrease of the surface energy, driving the system to a state where the liquid is confined to a set of spherical droplets. 
A sinusoidal distortion of the fluid cylinder of an initial radius $R_0$ is given by the radius dependence $r(z)=R_0^*+A\sin{qz}$, where $A$ is the distortion amplitude, $q$ is the wave number, and $R_0^*=R_0-A^2/4 R_0$ is the modified cylinder radius under the assumption of the volume conservation. The corresponding contribution of the surface energy is expressed by the formula:
\begin{equation}
  \Delta E=\frac{\pi  \lambda  \sigma }{2 R_0}(q^2 R_0^2 - 1) A^2.
  \end{equation}
While the surface energy is quadratically dependent on $A$, the undistorted state's stability hinges on the prefactor $\frac{\pi  \lambda  \sigma }{2 R_0}(q^2 R_0^2 - 1)$. Exceeding an aspect ratio of $\pi$ triggers instability in the lowest mode $q=2 \pi n/L$ with $n=1$, which is referred to as the Rayleigh-Plateau limit.

Distortions in the form of surface undulations result in the deformation of the nematic director within the filament. Such deformations are driven by the strong anchoring at the liquid/air interface. The distortion of the director field in bulk is controlled by the nematic elasticity and determined by the splay, twist and bend elastic constants ($K_{11}$, $K_{22}$, $K_{33}$, respectively)~\cite{deGennes:1995vg}. Assuming planar anchoring at the filament's interface and parametrising the nematic director using the angle $u(r,z)$ as $\mathbf{n}=(\cos(u), \sin(u))$, we obtain in \emph{one-constant} approximation ($K_{11}=K_{22}=K_{33}=K$) $u(r,z)$ from a solution of the Laplace equation (Eq.~\ref{eq:laplace}).

In the ferroelectric nematic with $\mathbf{P}_\mathrm{s}\propto \mathbf{n}$, surface undulations are accompanied by the splay of the director and, as a result, by undulations of the spontaneous polarisation. Since the polarisation splay results in the bound electric charge density $\rho_{\mathrm{b}}=-\nabla\cdot\mathbf{P}_\mathrm{s}$, the surface undulations will cost electrostatic energy. As Bellini et al.~\cite{Caimi.2023t5j} demonstrated the ferroelectric nematic phase confined in microchannels exhibits a unique property known as electric "super-screening." This phenomenon enables the polarisation in the N$_{\textrm{F}}$ to be restricted to a designated channel or filament, guiding the electric field. Specifically, any transverse electric field that exists with respect to the filament axis is rapidly compensated for by the induced bound electric charges. This compensation process leads to a confinement of the electric field of the polarised filament within the filament itself.
Electric displacement in a ferroelectric is given by the equation $\mathbf{D}=\varepsilon_0\varepsilon \mathbf{E}+\mathbf{D}_{\mathrm{r0}}$, where $\mathbf{E}$ is the electric field and $\mathbf{D}_{\mathrm{r0}}$ is the remanent displacement, equivalent to the spontaneous polarization $\mathbf{P}_{\mathrm{s}}$. Assuming no free charges, introducing the electric potential $\varphi$ with $\mathbf{E}=-\nabla\varphi$  we can derive $\varphi$ from the Poisson equation  
$\varepsilon_0\varepsilon\nabla^2\varphi=\nabla\cdot\mathbf{D}_{\mathrm{r0}}$,
where the term $\nabla\cdot\mathbf{D}_{\mathrm{r0}}
=-\rho_\mathrm{b}$ represents the density of the bound charges $\rho_\mathrm{b}$ and is determined by the deformation of the polar director field as $\mathbf{D}_{\mathrm{r0}}(u)=D_{\mathrm{r0}}[\cos(u), \sin(u)]$. The solutions of Eqs.~\ref{eq:poisson} and \ref {eq:laplace} provides required potential: 
\begin{equation}
  \nabla^2\varphi=\frac{1}{\varepsilon_0\varepsilon}\nabla\cdot\mathbf{D}_{\mathrm{r0}}(u).
  \label{eq:poisson}
\end{equation}
\begin{equation}
  \nabla^2 u=0.
  \label{eq:laplace}
\end{equation}
The electrostatic energy $ E_{\textrm{el}}$ is determined by the integration over the volume of the filament:
\begin{equation}
  E_{\textrm{el}}=\int \rho_{\textrm{b}}\varphi\textrm{d}v
\end{equation}

\begin{figure}[h!]
\centering
\includegraphics[width=\columnwidth]{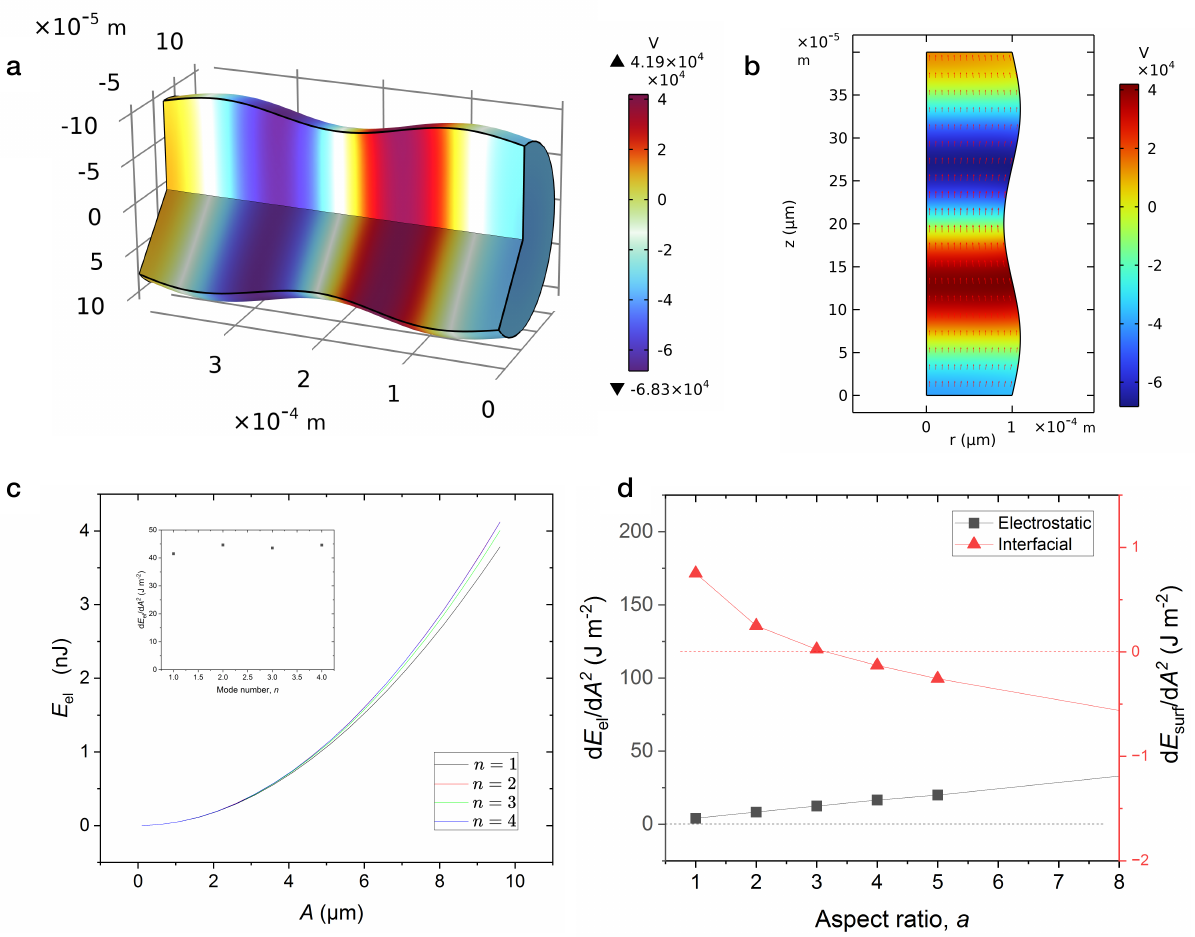}
\caption{Numerical simulations of electrostatic energy of a deformed filament: \textbf{a} Electric potential $\varphi$ in an exemplary filament with $L=\qty{40}{\micro\meter}$, initial radius $r_0=\qty{100}{\micro\meter}$ undulated with sine waveform (amplitude $A=\qty{10}{\micro\meter}$, mode number $n=3$. \textbf{b} Electric potential $\varphi(r,z)$ superimposed with the vector diagram of $\mathbf{D}_{\mathrm{r0}}(r,z)$ in a crosssection of the filament in \textbf{a}. \textbf{c} Dependence of the electrostatic energy on the undulation amplitude $A$ for the first four modes ($L=\qty{2000}{\micro\meter}$, $r_0=\qty{100}{\micro\meter}$); the inset shows the slope $\mathrm{d}E_\mathrm{el}/\mathrm{d}A^2$. \textbf{d} The dependence of the $\mathrm{d}E_\mathrm{el}/\mathrm{d}A^2$ and $\mathrm{d}E_\mathrm{surf}/\mathrm{d}A^2$ on the aspect ratio $a$ for a filament with the initial radius $r_0=\qty{100}{\micro\meter}$.
 }
\label{fig:model}
\end{figure}

Numerical solutions of the Eqs.~\ref{eq:poisson} and \ref{eq:laplace} for an exemplary filament of the length $L=\qty{40}{\micro\meter}$ and the initial radius $r_0=\qty{100}{\micro\meter}$ are shown in Fig.~\ref{fig:model}a. The filament's surface is undulated with $r(z)=r_0^*+A \sin{ 2\pi n z/L}$, $n=3$. The potential $\varphi(r,z)$ has an axial symmetry and exhibits a modulation as shown in the cross-section in Fig.~\ref{fig:model}b together with the residual displacement $\mathbf{D}_{\mathrm{r0}}(r,z)$. 

By keeping the modulation amplitude below 10\% of $r_0$, the electrostatic energy increases quadratically with $A$ and is mostly unaffected by the mode number $n$ (Fig.~\ref{fig:model}c). The slope $s_{\mathrm{el}}=\mathrm{d}E_\mathrm{el}/\mathrm{d}A^2$ only slightly increases with higher wave numbers. As a result, thus it is adequate to only consider the lowest mode with $n=1$.
With the elastic constant $K$ in the range of piconewtons, the elastic energy given by the energy density term $f_\mathrm{elastic}=K\nabla^2 u$ is several orders of magnitude smaller than the surface~\cite{Stannarius_PRE_2005} and electrostatic contributions.

When the length of a filament increases while the radius $r_0$ is fixed, the slope $s_{\mathrm{el}}$ linearly increases with the aspect ratio $a=L/r_0$ (black curve in Fig.~\ref{fig:model}d). Conversely, the surface energy decreases as the aspect ratio increases, and the slope $s_{\mathrm{surf}}=\mathrm{d}E_\mathrm{surf}/\mathrm{d}A^2$ changes sign from positive to negative, ultimately making long filaments unstable (red curve in Fig.~\ref{fig:model}d).

However, the electrostatic contribution exceeds the surface one by almost an order of magnitude, which explains why the breaking instability is suppressed. A drastic slowing down of the thinning dynamics in filaments prepared in an electric field is another confirmation that the stabilisation is of electrostatic origin. Indeed the field-guiding property of the  NF filament~\cite{Caimi.2023t5j} results in an electric field induced by the polarisation charges at the junctions to the menisci. The filament acts as a capacitor, and the electric field couples to the spontaneous polarisation within the fibre. 
Nonetheless, this stabilisation mechanism does not prevent the filament's thinning from occurring due to the transversal tension and material transport into the menisci. Another structural feature contributing to filament stabilisation is required to sustain the transversal stresses. One possible mechanism is the conjectured spontaneous polarisation splay creating a smectic or columnar-type superstructure that can balance the transversal stresses. Such splay is of flexoelectric nature and is the consequence of the symmetry of N$_{\textrm{F}}$. The linear splay term is proposed in the theoretical models of the N$_{\textrm{F}}$ phase~\cite{Pleiner.1989,Kats.2021,Mertelj.2018,Sebastian.2023g2e}. The experimental observations of the spontaneous splay were suggested by Mertelj and Sebasti\'an in~\cite{Sebastian.2023g2e} using photo-patterned substrates. A detailed theoretical description accounting for the conductivity, free charges, and polarisation fluctuations is required to accurately describe the stability of the N$_{\textrm{F}}$ filaments.

\section{Conclusion}\label{sec13}

In summary, we demonstrated that the recently discovered ferroelectric nematic phase can form freely suspended filaments in a addition to freely suspended films. The filaments exhibit a remarkable efficiency of the optical second harmonic conversion and their stability is strongly dependent on the external electric field.

Although those features are rather typical for the (modulated) smectic and columnar phases, they are yet observed in the nematic phase suggesting the suppression of the surface fluctuations and probably the presence of the polarisation-driven secondary structure such as a periodic director modulation.

\section{Methods}\label{sec11}

\textbf{Materials.} We investigated two liquid crystal compounds both synthesised and provided by Merck Electronic KGaA: Compound \textbf{1} is 4-((4'-butyl-2',3,5,6'-tetrafluoro[1,1'-biphenyl]-4-yl)difluoromethoxy)-2,6-difluorobenzonitrile
exhibiting a monotropic ferroelectric nematic phase near room temperature. The phase diagram is isotropic (\qty{19.6}{\celsius} N$_{\rm{F}}$) \qty{44.0}{\celsius} crystal.

Material \textbf{2} is a mixture exhibiting the nematic (N), an intermediate unclassified M phase, and the ferroelectric nematic (NF) phases: isotropic \qty{87.0}{\celsius} N \qty{57.0}{\celsius} M \qty{46.0}{\celsius} \qty{-43.0}{\celsius} crystal.\\

\noindent \textbf{Polarised light microscopy.} Optical studies were made using a polarised light microscope AxioImager A.1 (Carl Zeiss GmbH, Germany) equipped with a heating stage (Instec, USA). Using a custom-made pulling device, the filaments were pulled mechanically in a heating stage. Birefringence measurements were done using the Berek tilting compensator. \\

\noindent \textbf{Nonlinear optical microscopy.} The generation of the optical second harmonic (SHG) was measured using the multiphoton laser of a confocal microscope (Leica TCS SP8, CLSM). A tunable IR laser ($\lambda_\mathrm{ex}=\qty{880}{\nano\meter}$) was used as a fundamental light beam, and the emission was acquired at $\lambda_\mathrm{em}=\qty{880}{\nano\meter}$. As a reference, we used a \qty{50}{\micro\meter} $\alpha$-quartz crystal plate (z-cut).\\

\begin{figure}[h!]
\centering
\includegraphics[width=0.5\columnwidth]{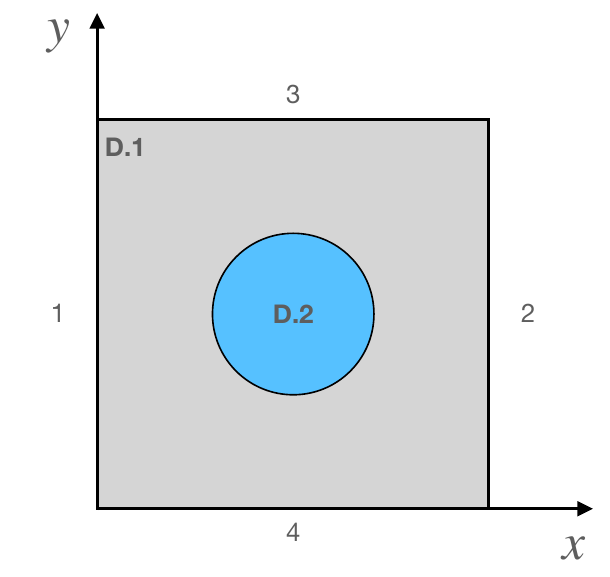}
\caption{Simulation domain for calculation of the SHG in a cylindrical filament. The simulation domain consists of the the inner SHG-active circular domain \textbf{D.2}, and the outer SHG-inactive domain \textbf{D.1}. the numbers designate the domain boundaries.  
 }
\label{fig:model_geometry}
\end{figure}

\noindent \textbf{Data analysis and simulations.} Data analysis and numerical simulations were made using Matlab (Mathworks) and Comsol software. To simulate SHG in filaments, we solved coupled equations for the fundamental and SH electromagnetic waves in the frequency domain. The geometrical domain is composed of an outer rectangular domain \textbf{D.1} and the inner cylinder with a circular cross-section in the $XY$-plane (domain \textbf{D.2}) as shown in Fig.~\ref{fig:model_geometry}. The cylinder's radius was varied from $0.2\lambda$ to $15\lambda$ to simulate the radius dependence of the SHG signal.

We considered a plane wave polarised along the cylinder axis propagating in $x$-direction: $\mathbf{E}(x, y, z)=\tilde{\mathbf{E}}(x, y) e^{-i k_x x}$.
The fundamental and the SH waves satisfy the wave equation: 
\begin{equation}
    \nabla \times \mu_r^{-1}(\nabla \times \mathbf{E}_l)-k_0^2\epsilon_r \mathbf{E}_l=\omega^2 \mu_0  \mathbf{P},
\end{equation}
where $l=f,s$ for the fundamental and the SH waves, respectively. The polarisation contains a sum of linear and non-linear contributions:
\begin{equation}
    \mathbf{P}=\epsilon_0\left(\epsilon_{\mathrm{r}}-1\right) \mathbf{E}_l+\mathbf{P}^{\mathrm{NL}},
\end{equation}
\noindent In the case of $z$-polarised beam, the nonlinear polarisation in the domain \textbf{D.2} is determined by the $d_{33}$ coefficient: 

\begin{equation}
  P^{\mathrm{NL}}_z=2\epsilon_0 d_{33} E_{fz}(\omega)^2
\end{equation}

At the domain \textbf{D.1} boundaries 1 and 2 satisfy scattering boundary conditions for the fundamental wave:
\begin{equation}
  \mathbf{n} \times(\nabla \times \mathbf{E}_l)-j k \mathbf{n} \times(\mathbf{E}_l \times \mathbf{n})=s
  \end{equation}
with $s=-\mathbf{n} \times\left(\mathbf{E}_0 \times\left(j k\left(\mathbf{n}-\mathbf{k}_{\mathrm{d} r}\right)\right)\right) e^{-j k \mathbf{k}_{\mathrm{dr}} \cdot \mathbf{r}}$ for boundary 1, and $s=0$ for boundary 2, and $\mathbf{n}$ is the normal to the boundary. The same conditions are satisfied for the SH wave with $s=0$ at both boundaries. The conditions for the inner boundary between the domains \textbf{D.1} and \textbf{D.2} are continuous.

\backmatter

%\bmhead{Supplementary information}
%
%If your article has accompanying supplementary file/s please state so here. 
%
%Authors reporting data from electrophoretic gels and blots should supply the full unprocessed scans for key as part of their Supplementary information. This may be requested by the editorial team/s if it is missing.
%
%Please refer to Journal-level guidance for any specific requirements.

\bmhead{Acknowledgments}

The authors would like to thank Antal J\'akli, Joseph Maclennan, Michail Osipov and Tommaso Bellini for fruitful discussions. N$_{\textrm{F}}$ and their electromechanical properties were also demonstrated by A. J\'akli at European Liquid Crystal Conference in Rende (Italy)~\cite{JakliECLC2023}.

The authors acknowledge the financial support of Deutsche Forschungsgemeinschaft (Project ER 467/8-3).

\section*{Declarations}

There are no conflicts to declare.

\end{document}